\documentclass{article}
\usepackage{amssymb}
\usepackage{epsf}
\usepackage{graphicx}

\newcommand{\postscript}[2]
{\setlength{\epsfxsize}{#2\hsize}\centerline{\epsfbox{#1}}}
\input{tcilatex}

\begin{document}

\author{Helio V. Fagundes\thanks{%
E-mail: helio@ift.unesp.br}\quad and Evelise Gausmann\thanks{%
E-mail: gausmann@ift.unesp.br} \\
\textit{Instituto de F\'{i}sica Te\'{o}rica, Universidade Estadual Paulista,}%
\\
\textit{\ Rua Pamplona, 145, S\~{a}o Paulo, SP 01405-900, Brazil}}
\title{Cosmic crystallography with a pullback }
\date{31 May 1999 }
\maketitle

\begin{abstract}
We present a modified version of the cosmic crystallography method,
especially useful for testing closed models of negative spatial curvature.
The images of clusters of galaxies in simulated catalogs are ``pulled back''
to the fundamental domain before the set of distances is calculated.

\end{abstract}

\section{Introduction}

The method of cosmic crystallography was devised by Lehoucq,
Lachi\`{e}ze-Rey, and Luminet \cite{LeLaLu} to investigate the global
spatial topology of the universe. See also \cite{FG98}.

\qquad Succinctly (see \cite{LaLu} for details), while the spatial sections
of a Friedmann-Lema\^{i}tre-Robertson-Walker (FLRW) cosmological model are
usually taken to be one of the simply connected spaces of constant curvature
(spherical space $S^{3},$ Euclidean space $E^{3},$ and hyperbolic space $%
H^{3})$, they can more generally be represented by a quotient manifold $M=%
\tilde{M}/\Gamma $, where $\ \tilde{M}$ is one of the mentioned spaces and $%
\Gamma $ is a discrete group of isometries (or rigid motions) acting freely
and properly discontinuously on $\tilde{M}$. In practice $M$ is described by
a \textit{Dirichlet domain }or \textit{fundamental polyhedron} ($FP$) in $%
\tilde{M},$ with faces pairwise identified through the action of the
elements of $\Gamma $; the latter is said to \textit{tesselate} $\tilde{M}$
into cells which are replicas of $FP$, so that $\tilde{M}=\Gamma (FP).$
Mathematically $\tilde{M}$ is the universal covering space of $M$, while
physically it is the locus of repeated images of sources in $M,$ one for
each cell.

\qquad The method of \cite{LeLaLu} consists of plotting the distances
between cosmic images of clusters of galaxies vs. the frequency of
occurrence of each of these distances. If real space turns out to be such a
manifold, and $\Gamma $ contains Clifford translations (cf. \cite{Gomero}, 
\cite{Uzan98}), then we may expect to see neat peaks or spikes in this plot,
and their pattern would be related to the topology of cosmic space.

\qquad In the case of hyperbolic space, the only Clifford translation in $%
\Gamma $ is the trivial motion \cite{Wolf}. Therefore the original
crystallography method may reveal very little, or nothing, of the global
topology; cf. \cite{Gomero}, \cite{Uzan98}, \cite{ccchu}. Some modified
forms of the method have been proposed for application to hyperbolic models;
see \cite{ccchu}, \cite{Gomero}, \cite{Uzan98}, \cite{Uzan99}. All of these,
as well as more general proposals based on other principles - see, for
example, \cite{CSS}, \cite{LuRo} and references there - are dependent on 
much expected, needed improvements on the observational side, and will 
certainly be found
to be complementary to each other. Here we present still another variant of
the crystallography idea, which may be particularly useful if it becomes
known that space is a hyperbolic manifold. Namely, we assume space to be a
definite hyperbolic manifold, with a fixed $FP$ in a fixed orientation in
astronomical space. The position of the observer inside $FP$ does not
influence the result, so one may work as if he or she were located at the
center (or \textit{basepoint)} of $FP$.\ 

The original crystallography scheme relied on the elements of $\Gamma $
bringing a source's position in $FP$ to the positions of its repeated images
in other cells. For hyperbolic manifolds a given action $\gamma \in \Gamma $
on points $p\in FP$ depends on $p$, and thus we do not get the wealth of
equal distances that make up the neat peaks in the case of Clifford
translations.

In the present variation of the method, we pull each image back to its
pre-image's position in $FP.$ If space is really $H^{3}/\Gamma ,$ with the
observer's position at the center of $FP$, then the distribution of
distances between the sources and the pulled back images will strongly peak
at zero distance - more precisely, near zero, because of little known facts
like evolution and peculiar velocities, and data inaccuracies. If real space
is rotated with respect to the assumed manifold, $FP$ will also be rotated,
and the pullback operation does not bring an image to its real pre-image's
position. So the neat peak at zero distance is destroyed; if the rotation is
small ($\lesssim 5^{\circ }$) a less sharp peak is still visible near zero
distance, but it quickly disappears as the angle of rotation increases.
Therefore one might have to check about a thousand orientations of $FP$, for
each candidate manifold in order to find the significant peak near zero
distance.

\section{The simulated catalogs}

\qquad\ We shall be working with the spacetime metric of FLRW hyperbolic
model, 
\[
ds^{2}=a^{2}(\eta )(d\eta ^{2}-d\lambda ^{2})\,, 
\]

where $a(\eta )$ is the expansion factor or curvature radius, and 
\[
d\lambda ^{2}=d\chi ^{2}+\sinh ^{2}\chi (d\theta ^{2}+\sin ^{2}\theta d\phi
^{2}) 
\]
is the standard or normalized metric of $H^{3}$ - cf.\cite{CTF}. We assume
for the cosmological parameters the values $\Omega _{0}=0.3,$ $H_{0}=65$
km\thinspace s$^{-1}$Mpc$^{-1},$ and $\Lambda =0.$ The present value of the
curvature radius is then $a(\eta _{0})=5512.62$ Mpc.

\qquad Our computer simulated catalogs are similar to those in \cite{ccchu}.
There is an improvement in the making of tables of pseudo-random points: The
volume element in hyperbolic space is $dV=\sinh ^{2}\chi \sin \theta d\chi
\,d\theta \,d\phi $; if we define new coordinates $u(\chi )=(\sinh \chi
\cosh \chi -\chi )/2$, $v(\theta )=\cos ^{-1}\theta ,$ and $\phi =\phi ,$ we
get $dV=du\,dv\,d\phi ,$ so that the probability density of points in ($%
u,v,\phi )$-space is uniform, and we need not weight the distribution of
random values for these coordinates.

\qquad We did the simulations for two compact hyperbolic models, both with a
regular isosahedron as $FP$ but different groups $\Gamma .$ They are the
first and second manifolds in Table 1 in \cite{Best}, and appear in the \
`closed census' of \textit{SnapPea }\cite{JW} as $v2051(+3,2)$ and $%
v2293(+3,2)$, respectively. The results are quite similar for both
manifolds; here we will only report those for $v2293$. We took from \cite
{ccchu} the set of 92 neighboring copies of $FP$. These 93 cells (including
the original $FP$) completely cover a ball of radius $\overline{\chi }%
=2.33947$ in $H^{3}.$ If we take $Z=1300$ for the redshift of the last
scattering surface (SLS) we get for the latter's normalized radius $\chi
_{SLS}=2.33520<\overline{\chi }, $ hence the presently observable universe
fits within the described 93-cell region.

\qquad To build simulated catalogs for the compact models we first created \
100 random points inside the ball with the radius $\chi _{out}=1.38257$ of
the sphere circumscribing the icosahedron. We then excluded those points
outside $FP,$ and took the $29$ points remaining inside as sources. From the
latter a datafile of 897 potential images within the $\chi _{SLS}$-ball
(including the sources themselves) was created, using the 93 elements of $%
\Gamma $ that cover this ball.

\qquad However, since we intended to displace the observer to position $%
(0.1, 0, 0)$ in Klein coordinates, which is at a distance $\delta =\tanh
^{-1}0.1 $ from the center, we took the catalogs' radii to be $\chi _{\max
}=\chi _{SLS}-\delta =2.23486.$

\qquad Two catalogs were then prepared from the $\chi _{SLS}$-ball datafile,
both with radius $\chi _{\max }$ and the orientation of the \textit{SnapPea}%
's coordinates, centered on the observer at (0,0,0) in one catalog and at
(0.1, 0, 0) in the other.

\section{The pullback crystallography}

\subsection{Observer at basepoint}

\qquad Assuming that our $v2293$ manifold represents cosmic space, with the
same orientation and basepoint as given by \textit{SnapPea}, our first
simulation places the observer at the basepoint, which is the center of the
icosahedron.

\qquad For all images $q$ in the catalog with radius $\chi _{\max }$, we
find their pre-images in $FP$, that is, we find $p=\gamma ^{-1}q$, $\gamma
\in \Gamma $. The computer procedure for this process was essentially
written by Weeks\textsc{\ }\cite{JWpc}, in the context of \textit{SnapPea};
it is based on the very definition of a Dirichlet domain.

\qquad Let $S$ be the set of sources in $FP$, and $n(p)$ be the number of
images in the catalog with pre-image $p\in S$. The pullback process takes
all of them to $p.$ When the distances between the sources and the pulled
back images are calculated, there will be $\stackunder{p\in S}{\tsum }n(p)$ 
null distances, hence a strong peak at zero. The counting for other
distances follows a pattern similar to those of the failed attempts to do
the standard crystallography with hyperbolic manifolds - cf. \cite{ccchu}
and references there. The result is shown if Fig.\thinspace 1a, with
percentage of occurrences plotted versus distances, the latter in bins of
100 Mpc.

\qquad In Fig. 1b we show the result of pulling back the images of a random
distribution of sources in the open FLRW model. A catalog with the same
radius was assumed, and the sources were pulled back with the operation of
the same group $\Gamma $ as above. Fig. 1c shows the differences between the
previous two plots.

\begin{figure}
\postscript{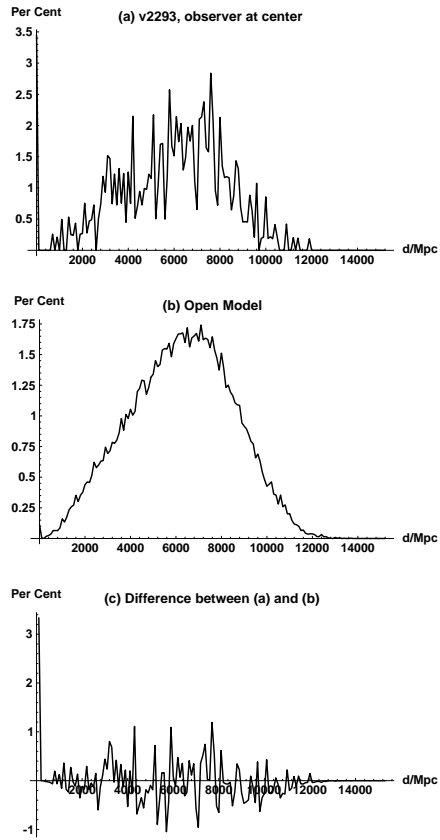}{0.7}
\caption{The results of crystallography with pullback for a simulated catalog
of clusters of galaxies. Figure \textit{a }corresponds to the multiply
connected model, with the observer at the center of the fundamental region,
\textit{b} to a simply connected universe with the same physical parameters
as \textit{a, }and\textit{\ c }to the difference between \textit{a }and%
\textit{\ b.} }
\end{figure}

\subsection{Observer displaced from basepoint}

\qquad Moving the observer from $(0,0,0)$ to $(0.1,0,0)$, we built a new
catalog, with images from the $\chi _{SLS}$-ball with distances up to $\chi
_{\max }$ from the new center. Then we proceeded as above, and obtained
practically the same results. This is as expected, since the set $S$ of
sources and the pulled back images are the same, only with a new set of
frequencies $n^{\prime }(p)$, whose sum is approximately the same as before.

\subsection{Manifold in a different orientation}

\qquad Our next step is to rotate $FP$. Suppose our manifold with its
sources and repeated images represents the real distribution of cosmic
images, with the observer at $(0,0,0).$ Also suppose we do not know that the
orientation of this manifold in space is that implied by the coordinates
used in the \textit{SnapPea }census. We want to try the pullback process
with several rotated versions of $FP$, to see if the characteristic peaks
near zero distance are still present.

\qquad If we represent $H^{3}$ as the upper branch of hyperboloid $%
X_{0}^{2}-X_{1}^{2}-X_{2}^{2}-X_{3}^{2}=1$ in Minkowski space, then the
rigid motions in $H^{3}$, like the elements $\gamma \in \Gamma ,$ are
represented by $4\times 4$ Lorentz transformation matrices (see, for
example, \cite{QGAII}). A rotation with Euler angles $\phi ,\theta ,\psi $
corresponds to a matrix $R_{\mu \nu }$ $(\mu ,\nu =0-3),$ where $R_{0\nu
}=R_{\nu 0}=\delta _{\nu 0}$ and $R_{ij}=R_{ij}(\phi ,\theta ,\psi )$ is the 
$3\times 3$ rotation matrix, as given in \cite{Goldstein}. Let the
face-pairing generators of $\Gamma $ in \textit{SnapPea }be $\gamma _{k},$ $%
k=1-20;$ then the generators will be $\gamma _{k}^{\prime }=R\gamma
_{k}R^{-1}$ for the rotated manifold. The result is that if the images are
pulled into the rotated $FP$, using the new generators, they no longer
coincide with their pre-images in $FP$, with the exception of eventual
images of our own Galaxy.

\qquad Because of this the peak near zero distance tends to quickly
disappear with increasing angle of rotation. In Figs.\thinspace 2a-c we show
the difference plots for angles $(\phi ,\theta ,\psi )=(2^{\circ },0,0),$ $%
(0,5^{\circ },0),$ and $(150^{\circ },100^{\circ },60^{\circ })$,
respectively. The plot for a pulled back random distribution has been
subtracted out, as in Fig.\thinspace 1-c. The peak near zero is quite
visible in the first, is less sharp in the second, and does not appear at
all in the third case.

\qquad On the other hand, away from zero distance, these plots resemble the
difference plot in \cite{ccchu}, whose topological significance is still
uncertain - cf. \cite{UzanTexas}.

\begin{figure}
\postscript{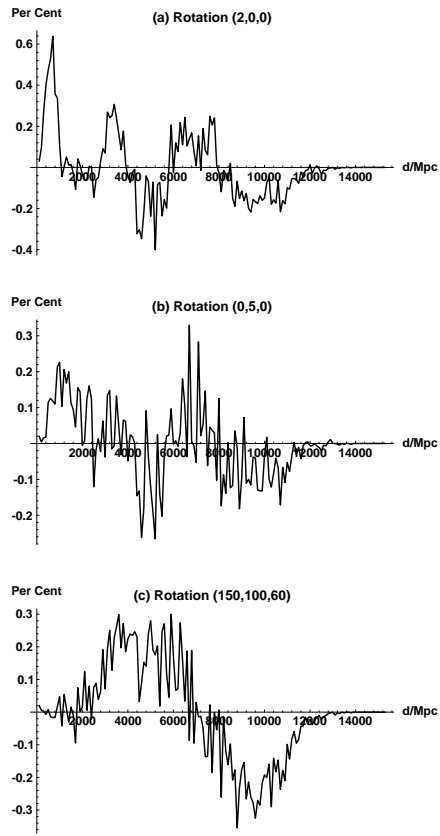}{0.7}
\caption{The difference between the plots for three models with the same 
fundamental polyhedron in different orientations in space, and that for the 
simply connected universe of Fig. 2b.}
\end{figure}

\section{Conclusion}

\qquad As suggested in Sec. 1, this method would be most useful if we
previously knew that space could be a given manifold. This emphasizes our
belief that the search for the true cosmic topology will not be easy or
straightforward, given the usual inaccuracies and fragmentary nature of
observational data. The various proposals that have appeared (cf. \cite{LuRo}%
) and will continue to appear should all contribute to this much desired
discovery, its confirmation and development, and its role in the building of
a new, richer, and truer cosmological picture.
\begin{verbatim}
 
\end{verbatim}

\qquad We are grateful to Jeff Weeks for the routine program \cite{JWpc}. EG
thanks Conselho Nacional de Desenvolvimento Cient\'{i}fico e Tecnol\'{o}gico
(CNPq - Brazil) for a doctorate scholarship. HVF thanks G. Matsas and R.
Rosenfeld for conversations on computer simulation.

\bigskip

\end{document}